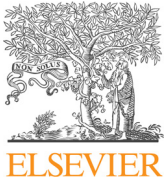
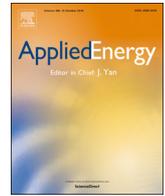

# Probabilistic duck curve in high PV penetration power system: Concept, modeling, and empirical analysis in China

Qingchun Hou[a], Ning Zhang[a,*], Ershun Du[a], Miao Miao[b], Fei Peng[c], Chongqing Kang[a,*]

[a] Department of Electrical Engineering, Tsinghua University, Beijing 100084, China
[b] School of Electrical and Electronic Engineering, Huazhong University of Science and Technology, Wuhan, China
[c] Economic Research Institute, State Grid Qinghai Electric Power Company, Xining, China

## HIGHLIGHTS

- The concepts of probabilistic duck curve and probabilistic ramp curve are proposed.
- The probabilistic duck curve is modeled considering dependencies among PV and loads.
- Probabilistic duck curve is used for flexible resource planning.
- Empirical analysis is conducted on actual high PV penetration power system in China.

## GRAPHICAL ABSTRACT

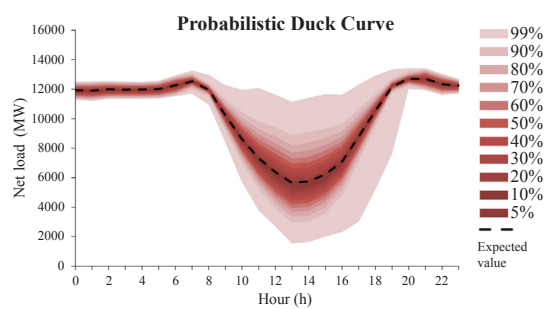

## ARTICLE INFO

## ABSTRACT

The high penetration of photovoltaic (PV) is reshaping the electricity net-load curve and has a significant impact on power system operation and planning. The concept of duck curve is widely used to describe the timing imbalance between peak demand and PV generation. The traditional duck curve is deterministic and only shows a single extreme or typical scenario during a day. Thus, it cannot capture both the probability of that scenario and the uncertainty of PV generation and loads. These weaknesses limit the application of the duck curve on power system planning under high PV penetration. To address this issue, the novel concepts of probabilistic duck curve (PDC) and probabilistic ramp curve (PRC) are proposed to accurately model the uncertainty and variability of electricity net load and ramp under high PV penetration. An efficient method is presented for modeling PDC and PRC using kernel density estimation, copula function, and dependent discrete convolution. Several indices are designed to quantify the characteristics of the PDC and PRC. For the application, we demonstrate how the PDC and PRC will benefit flexible resource planning. Finally, an empirical study on the Qinghai provincial power system of China validates the effectiveness of the presented method. The results of PDC and PRC intuitively illustrate that the ramp demand and the valley of net load face considerable uncertainty under high PV penetration. The results of flexible resource planning indicate that retrofitting coal-fired units has remarkable performance on enhancing the power system flexibility in Qinghai. In average, reducing the minimal output of coal-fired units by 1 MW will increase PV accommodation by over 4 MWh each day.

## Nomenclature

| | |
|---|---|
| $P_t$, $P_t^d$, $P_t^{PV}$ | net load, total load and total PV generation at period $t$, respectively |
| $f(\cdot)$ | PDF |
| $\otimes$ | convolution sign of PDF considering dependency |
| $\ominus$ | subtraction sign of DPS considering dependency |
| $\oplus$ | sum sign of DPS considering dependency |
| $R_t$ | ramp demand of the net load at period $t$ |
| $f_h(\cdot)$ | Kernel density estimation |
| $K(\cdot)$ | Kernel function |
| $F(\cdot)$ | CDF |
| $F_{n,t}^{PV}(\cdot)$, | marginal CDF of the $n$th PV generation at period $t$ |
| $F_{m,t}^{load}(\cdot)$ | marginal CDF of the $m$th nodal load at period $t$ |
| $C(\cdot)$ | copula function |
| $C_t^{PV-PV}$ | Gaussian copula function between the two PV farms at period $t$ |
| $C_t^{load\_load}$ | Gaussian copula function between the two loads at period $t$ |
| $C_{m,t}^{PV\_load}$ | Gaussian copula function between total PV generation and load at period $t$ |
| $C_t^{netloadt+1\_netloadt}$ | Gaussian copula function between net load at time period $t$ and $t+1$ |
| $C^{pk\_vy}$ | Gaussian copula function between peak time and valley time of net load |
| $\rho$ | correlation parameter of copula function |
| $\tau$ | Kendall correlation |
| $\Phi(\cdot)$ | CDF of standard normal distribution |
| $\Sigma$ | correlation coefficient matrix |
| $\Delta p$ | fixed step for distribution discretization |
| $Q_{n,t}^{PV}$ | DPS of the $n$th PV generation at period $t$ |
| $Q_{m,t}^{load}$ | DPS of the $m$th nodal load at period $t$ |
| $Q_t^{TotalPV}$ | DPS of total PV generation at period $t$ |
| $Q_t^{Totalload}$ | DPS of total load at period $t$ |
| $Q_t^{PDC}$ | DPS of PDC at period $t$ |
| $Q_t^{PRC}$ | DPS of PRC at period $t$ |
| $N$, $M$ | number of PV farm and node, respectively |
| $P_t^{min}$, $R_t^{min}$ | minimal net load and minimal net load ramp at period $t$, respectively |
| $E_t^{netload}$ | expected values of net load at period $t$ |
| $E_t^{ramp}$ | expected values of net load ramp at period $t$ |
| $L_{\alpha\%}$ | $\alpha\%$ quantile of net load or net load ramp distribution |
| $CL_{\alpha\%}$ | $\alpha\%$ confidence level |
| $Q^{PR}$ | DPS of PTV |
| $S$ | probabilistic area |
| $\Delta S$ | marginal probabilistic area |
| $P_t^{MOU}$ | MOU at period $t$ |

The main notation used in this paper is provided in Nomenclature; other symbols are defined as required.

## 1. Introduction

Photovoltaic (PV) cost reduction and global warming issue are promoting rapid growth of PV capacity around the world [1,2]. The accumulative PV capacity in China reached 130 GW in 2017, ranking first in the world [3]. By 2020, China plans to achieve over 200 GW of PV. As the highest PV penetrated region in China, Qinghai province plans over 10 GW PV in 2020, accounting for approximately 77% of its peak load demand. Due to the inherent variability and uncertainty of PV generation, high PV penetration certainly will bring major challenges to power system operation and planning, especially the demand for flexible resources.

The California Independent System Operator (CAISO) and Hawaii utilities have grown accustomed to a high PV penetration in their power systems [4,5]. The concept of the duck curve was introduced to describe the impact of high PV penetration on power system active power balancing [6]. The duck curve is defined as the total actual electricity load curve minus the renewable energy generation, especially in a high PV penetration scenario. The duck curve should be balanced by flexible resources in the power system. Fig. 1 illustrates the changes in the duck curve of CAISO from 2012 to 2020 and shows increasing demand for peak regulation. The duck curve remains relatively stable at night, decreases rapidly during sunrise, reaches the bottom at noon, increases sharply during sunset and finally peaks in the evening. This indicates that the duck curve will place considerable peaking and ramping regulation stress on conventional dispatchable generators.

Many studies have been conducted to explore the impact of the duck curve on power system operation and planning. Some studies focus on how to accommodate the PV generation when the duck curve reaches the valley. A report from the National Renewable Energy Laboratory (NREL) examined the potential overgeneration risk based on the shape of the duck curve [7]. Schoenung and Keller focused on revenue potential for using PV overgeneration at noon for electrolysis to produce hydrogen for fuel cell electric vehicles and to accommodate more renewable energy [8]. Zhang et al. employed electric boilers for heat and pumped hydro for energy storage to enable greater renewable energy penetration in China [9]. Denholm and Margolis examined several options, such as reducing the minimal load on conventional generators, load shifting, and energy storage to increase the PV penetration beyond 20% [10]. Komušanac et al. studied the impact of high renewable energy penetration on the power system load and concluded that electricity generation from renewable energy could reach 70% in Croatia [11]. Chaudhary and Rizwan proposed an energy management system using pumped storage and advanced PV forecast method to accommodate PV generation [12].

There are several strategies to improve the power systems flexibility to mitigate the impact of the duck curve, including demand response, energy storage, and retrofitting thermal plants. CAISO deployed demand response to change the duck curve shape to ensure grid reliability [13]. Floch et al. proposed a distributed charging algorithm for plug-in electric vehicles to flatten the duck curve [14]. Sanandaji et al. utilized thermostatically controlled loads such as air conditioners and refrigeration units to provide fast regulation reserve service during the sunset [15]. Lazar comprehensively proposed ten specific strategies to mitigate the challenges of the duck curve, including demand response and energy storage [16]. Hassan and Janko investigated how to use tariff incentives and time-of-use electricity prices to encourage deployment of PV with batteries, respectively [17,18]. Ding et al. compared the economics of conventional plants, energy storage, and customer demand response in providing flexibility in China [19]. Segundo et al. performed a techno-economic analysis of PV curtailment and energy storage in a Zurich distribution grid. The results show that PV curtailment is more cost-effective but may be unacceptable [20]. Kumar et al. conducted a cost-benefit analysis of retrofitting coal-fired plants in improving power system flexibility and accommodating renewable energy. The results indicate that reducing the power plant minimal output is the most cost-efficient approach [21]. This approach may be more effective in coal-dominated power systems in China, because the average minimal output rate of coal-fired units in China's north region is approximately 50% (part of the new units can reach as low as 40%), which is much higher than those in Denmark and Germany (approximately 20%). Therefore, the flexibility of coal-fired units in China could be substantially improved.

The duck curve provides intuitive insights into the impact of high PV penetration, such as steep ramps and overgeneration risk. Therefore, the duck curve is usually used for planning flexible resources such as peak regulation generators, energy storage, and demand response. However, the duck curve has the following shortcomings: (1) it only shows a single extreme or typical net load scenario and needs to identify those scenarios in advance; (2) it is deterministic and does not contain uncertainty and probability information such as the possible net load range and the occurrence probability of the extreme or typical scenario; (3) it cannot capture the variability and uncertainty of net load ramp; (4) it cannot accurately quantify the possible curtailment of renewable energy; and (5) the decision-making based on the deterministic duck curve tends to overinvest in flexible resources to avoid the renewable energy curtailment and reliability problems, since the occurrence probability of the extreme scenario is ignored. To address these issues, the uncertainty of daily PV generation should be fully considered in flexible resource planning and the deterministic duck curve should be improved to contain the uncertainty and probability information.

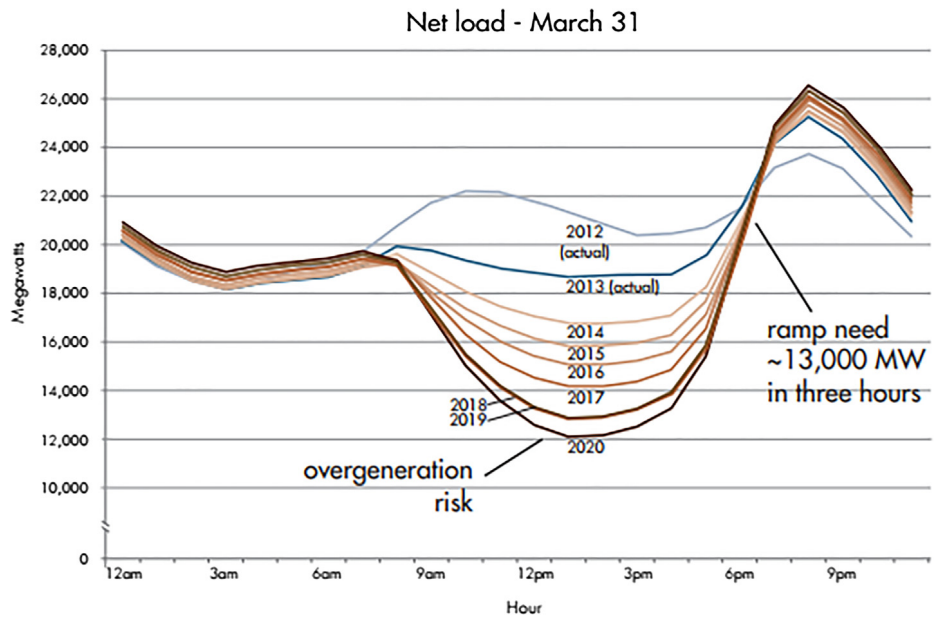

**Fig. 1.** Duck curve of California from 2012 to 2020 [1,6].

In this paper, the concept of probabilistic duck curve (PDC) is proposed. PDC provides the probabilistic distribution of electricity net load in each period during a day to consider the uncertainty of both PV generation and load (shown in Fig. 2). Based on the PDC, the concept of probabilistic ramp curve (PRC) is proposed to describe the ramping capacity requirement under high PV penetration from a probabilistic perspective. In addition, the PDC and PRC are visualized to intuitively identify the most representative and extreme scenarios. Since both PV generation and loads are affected by some common factors such as weather and temperature, there might be a complex dependence structure among PV generation and loads. The dependence structure is modeled using copula function [22] and dependent discrete convolution (DDC) [23] in the PDC and PRC. Furthermore, several indices are designed to quantify the characteristics of the PDC and PRC. In an empirical analysis, the PDC and PRC are visualized for the Qinghai power system in 2020 when the PV capacity reaches 10 GW and accounts for approximately 77% of its peak load demand. To further illustrate the application in actual power systems, the PDC is used in planning flexible resources such as retrofit of coal-fired units, energy storage and curtailment of PV generation.

The contributions of this paper are as follows:

(1) Propose the concepts of PDC and PRC to capture the uncertainty of the net load and ramp under high PV penetration.
(2) Provide the modeling method of PDC and PRC considering the complex dependencies among PV generation and loads.
(3) Provide a new point of view on the technical-economic analysis of flexibility resources for accommodating PV using the PDC and PRC. An empirical analysis is conducted on the Qinghai power system which has the highest PV penetration in China.

The remainder of this paper is organized as follows. Section 2 introduces the PDC and PRC methodology, including the concept, modeling method, and characteristic indices. An empirical study based on the Qinghai power system is conducted in Section 3. The application of the PDC and PRC for flexible resource planning is discussed in Section 4. Finally, conclusions and future works are in Section 5.

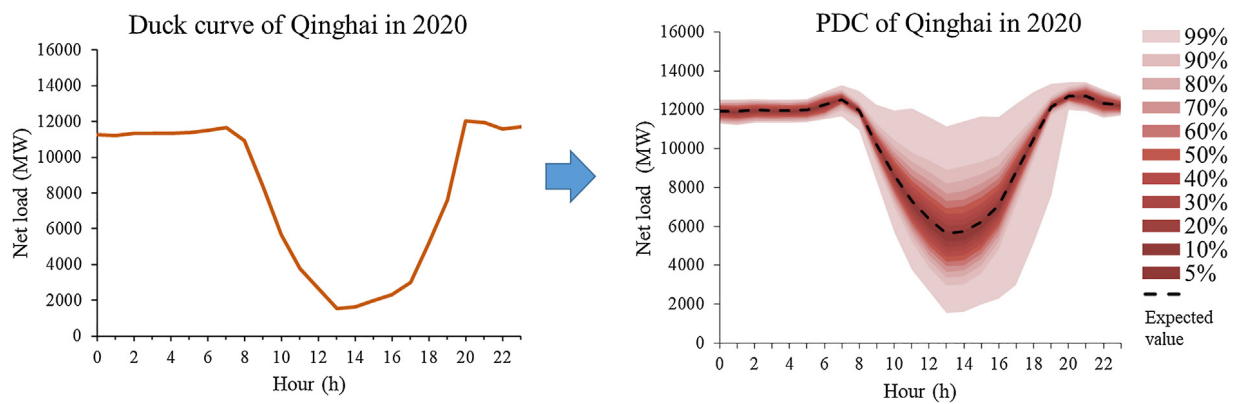

**Fig. 2.** Extending the duck curve to the probabilistic duck curve.

## 2. Methodology

### 2.1. Concepts of the probabilistic duck curve and probabilistic ramp curve

The PDC is defined as the set of probabilistic distributions of the net load for each period in a day considering the system load curve and PV generation. The probabilistic form of the duck curve considers the uncertainty of the load and the PV generation from a long-term perspective. Mathematically, the deterministic duck curve can be described as follows:

$$P_t = P_t^d - P_t^{PV} \tag{1}$$

where $t$ denotes the time period in a day, and $P_t$, $P_t^d$, and $P_t^{PV}$ represent the net load, total load and total PV generation at period $t$, respectively.

To present the duck curve in a probabilistic fashion, the distributions of both PV generation $f(P_t^d)$ and load $f(P_t^d)$ should be carefully modeled. In addition, in the process of subtraction, the dependencies should be considered. In this paper, we define the PDC as follows:

$$f(P_t) = f(P_t^d - P_t^{PV}) = f(P_t^d) \otimes f(-P_t^{PV}) \tag{2}$$

where $f(\cdot)$ is the probability density function (PDF), and $\otimes$ represents the convolution sign of PDF considering dependency. The calculation details of the convolution $\otimes$ will be introduced in the following part of this section.

The PDC contains both the uncertainty and dependence structure of the daily net load under high PV penetration. The PDC provides an intuitive approach to understanding how much and how frequently peak regulation is required and making it straightforward to recognize extreme or typical scenarios. Similarly, to illustrate the net load change between adjacent periods, the concept of the PRC is introduced as the set of probabilistic distributions of the net load ramp between adjacent periods. The PRC reveals the power system ramp demand in terms of both capacity and probability for each period in a day. Mathematically, this is expressed as follows:

$$f(R_t) = f(P_{t+1} - P_t) = f(P_{t+1}) \otimes f(-P_t) \tag{3}$$

where $R_t$ denotes the ramp of the net load at period $t$; $P_{t+1}$ and $P_t$ represent the net load at periods $t+1$ and $t$, respectively. It should be noted that the concepts of the PDC and PRC can readily be extended to power systems with high wind power penetration, which is beyond the scope of this paper.

The PDC and PRC reveal the increasing uncertainty of the net load and ramp under high PV penetration. Such uncertainty will fundamentally change the technical-economic analysis of flexible resources such as flexible generators, storage, and demand response. The PDC and PRC facilitate a more comprehensive technical-economic analysis and strategic planning for such flexible resources, which will be discussed in detail in Section 3.

### 2.2. Modeling method for the probabilistic duck curve and probabilistic ramp curve

According to the previous definition, modeling of both the PDC and PRC involves probabilistic calculation of load and PV generation, which cannot be implemented using arithmetic operation. Additionally, the various marginal distributions of PV generation and nodal loads and the complex dependencies among them add more challenges to PDC and PRC modeling. Thus, the key problems are how to achieve unified modeling of the marginal distribution of PV generation and loads, how to model the dependence structure among them, and how to calculate the probabilistic distribution effectively. Therefore, a kernel function, the copula function and dependent discrete convolution (DDC) are employed to facilitate separate modeling of marginal distributions and dependencies among PV generation and loads. A four-stage modeling method for the PDC and PRC is proposed, as shown in Fig. 3. First, the marginal distributions of load and PV generation are modeled using a nonparametric kernel density function. Second, the spatiotemporal dependencies among PV generation and loads are modeled using copula function. Third, the distributions of the total load and total PV generation are discretely calculated using DDC. Finally, the PDC and PRC are derived. The stages of PDC and PRC modeling will be presented in detail in the following parts of the Section 2.

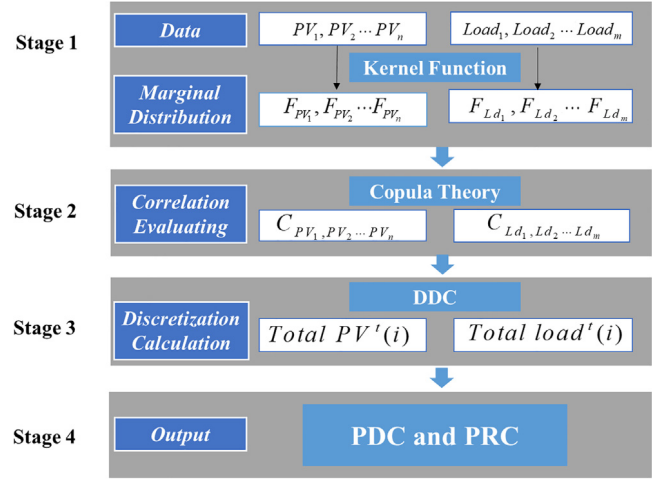

**Fig. 3.** Stages of PDC and PRC modeling.

#### 2.2.1. Marginal distributions of PV and load

The distributions of PV generation and load are usually modeled by parameter estimation functions such as beta, Weibull, and normal function. However, these well-defined distribution functions perform well only with a prior assumption that PV generation and load satisfy a certain distribution, which can hardly hold for various PV generation and load scenarios. Therefore, nonparametric kernel density estimation [24,25], which is suitable for arbitrary distribution, is used in this paper.

The kernel density estimation $\tilde{f}_h(x)$ of PV generation or load at point $x$ MW can be modeled as:

$$\tilde{f}_h(x) = \frac{1}{Lh} \sum_{i=1}^{L} K\left(\frac{x - X_i}{h}\right) \tag{4}$$

where $X_i$ is a sample from a PV generation or load dataset; $L$ is the number of total samples; $h$ denotes the bandwidth, that can be adjusted to avoid over or under fitting [26]; and $K(\cdot)$ represents the kernel function. The most commonly used Gaussian kernel function is expressed as follows:

$$K(u) = \frac{1}{\sqrt{2\pi}} \exp\left(-\frac{1}{2}u^2\right) \tag{5}$$

Looking into the form of the kernel function and kernel estimation, it actually assigns a weight to each sample. The farther the sample from the estimated point $x$, the smaller the weight. This result is reasonable because a sample far from the point $x$ has less impact on the point's distribution.

Substituting the PV generation and load datasets into Eqs. (4) and (5), we obtain the marginal cumulative distribution functions (CDFs) of the $n$th PV generation $F_{n,t}^{PV}(\cdot)$ and $m$th nodal load $F_{m,t}^{load}(\cdot)$ at period $t$.

#### 2.2.2. Dependence structure modeling with the copula function

To model the dependence structure among PV generation and loads, the copula function is further introduced after PV and load marginal distribution modeling. The copula function [22] is an effective method to model the dependence structure of stochastic variables, especially for non-Gaussian marginal distribution. We suppose that $x$ and $y$ are random variables representing the generation of two PV farms at period $t$. The corresponding marginal CDFs are $F_X$ and $F_Y$, respectively.

According to copula theory, the joint distribution $F_{XY}$ can be expressed as follows:

$$F_{XY}(x, y) = C(F_X(x), F_Y(y)) \qquad (6)$$

where $C(\cdot)$ denotes the copula function, which has a definition domain of $[0,1]^2$. Apparently, the copula function provides nearly the same information as the joint distribution.

Assuming that the dependence structure between the two PV farms corresponds to a Gaussian copula, the PDF of the Gaussian copula $C_t^{PV\_PV}$ is given as follows [27]:

$$C_t^{PV\_PV}(F_X, F_Y, \rho) = \frac{1}{\Sigma^{1/2}} \cdot \exp\left[-\frac{1}{2}\gamma'(\Sigma^{-1} - I)\gamma\right] \qquad (7)$$

where $\gamma'$ is the transpose of $\gamma$; $\gamma = [\Phi^{-1}(F_X), \Phi^{-1}(F_Y)]'$; and the matrix $\Sigma = \begin{bmatrix} 1 & \rho \\ \rho & 1 \end{bmatrix}$ denotes the correlation coefficient matrix between two PV farms. The correlation parameter $\rho$ is calculated as follows:

$$\rho = \sin\left(\frac{\pi\tau}{2}\right) \qquad (8)$$

where the Kendall correlation $\tau$ between the two PV farms is derived from ordered samples $(x_i, y_i)$ [28].

Substituting the PV generation and load datasets into Eqs. (6)–(8), the copula functions $C_t^{load\_load}$, $C_t^{PV\_load}$, and $C_t^{netload t+1\_netload t}$ are calculated to represent the dependence structure between load and load, total PV generation and total load, and net load at time periods $t$ and $t + 1$, respectively.

### 2.2.3. Dependent discrete convolution (DDC)

The marginal distributions and dependence structure of PV generation and loads are modeled with kernel density estimation and the copula functions, respectively. On this basis, the distribution of net load can be discretely calculated using the DDC theory [23]. DDC is an extension of the sequence operation theory, which provides an effective way to calculate the summation of dependent stochastic variables [29]. According to the DDC, the discrete probabilistic sequence (DPS) $s$, denoting the distribution of total PV generation $x + y$, can be modeled as follows:

$$s(i) = \sum_{i_a + i_b = i} C\left(\sum_{m=0}^{i_a} a(m), \sum_{n=0}^{i_b} b(n)\right) \cdot a(i_a) \cdot b(i_b), \quad i = 0, 1, \ldots, N_a + N_b \qquad (9)$$

where $s(i)$ is the $i$th element of $s$; $C(\cdot)$ is the PDF of the copula function; and $a$ and $b$ denote the DPSs of $F_X$ and $F_Y$, respectively. They can be derived by the following formulations:

$$a(i) = F_X(i\Delta p + \Delta p/2) - F_X(i\Delta p - \Delta p/2) \quad i = 0, 1, \ldots, N_a \qquad (10)$$

$$b(i) = F_Y(i\Delta p + \Delta p/2) - F_Y(i\Delta p - \Delta p/2) \quad i = 0, 1, \ldots, N_b \qquad (11)$$

where $\Delta p$ denotes the fixed step of discretization; $i$ represents the element serial number in the DPS; and $N_a + 1$ and $N_b + 1$ represent the length of $a$ and $b$, respectively.

Eq. (9) can be abbreviated as follows:

$$s(i) = a(i) \overset{C^{x\_y}}{\oplus} b(i) \qquad (12)$$

where $C^{x\_y}$ denotes the copula function between $x$ and $y$.

Because $x - y$ is equivalent to $x + (-y)$, the DPS of $x - y$ can be derived as follows:

$$d(i) = a(i) \overset{C^{x\_y}}{\ominus} b(i) = a(i) \overset{C^{x\_-y}}{\oplus} e(i) = a(i) \overset{C^{x\_y}(-\rho)}{\oplus} \hat{b}(i) \qquad (13)$$

where $d$ denotes the DPS of $x - y$; $e$ denotes the DPS of $- y$; $C^{x\_-y}$ denotes the copula function between $x$ and $- y$; $\hat{b}$ represents the reverse sequence of DPS $b$; and $C^{x\_y}(-\rho)$ represents the copula function between $x$ and $y$ with negative correlation coefficient $- \rho$.

According to Eqs. (10) and (11), the marginal distributions of PV

and load, $F_{n,t}^{PV}(\cdot)$ and $F_{m,t}^{load}(\cdot)$, can be discretized into DPSs of PV and load, $Q_{n,t}^{PV}$ and $Q_{m,t}^{load}$, respectively, with the same fixed step $\Delta p$. Substituting the DPSs $Q_{n,t}^{PV}$ and $Q_{m,t}^{load}$ into Eq. (12), the total PV generation DPS $Q_t^{TotalPV}$ and the total load DPS $Q_t^{Totalload}$ can then be derived as follows:

$$Q_t^{TotalPV}(i) = \sum_{k=1}^{N} \overset{C_t^{PV\_PV}}{\oplus} Q_{k,t}^{PV}(i) \qquad (14)$$

$$Q_t^{Totalload}(i) = \sum_{k=1}^{M} \overset{C_t^{load\_load}}{\oplus} Q_{k,t}^{load}(i) \qquad (15)$$

where $\sum \oplus$ represents the sum of the multiple dependent sequences [30]; $N$ and $M$ denote the number of PV farms and load nodes, respectively.

### 2.2.4. Probabilistic duck curve and probabilistic ramp curve

#### 1) Probabilistic duck curve

According to the PDC definition and Eq. (13), the DPS of PDC $Q_t^{PDC}$ can be derived from DDC between the total load DPS $Q_t^{Totalload}$ and total PV generation DPS $Q_t^{TotalPV}$:

$$Q_t^{PDC}(i) = Q_t^{Totalload}(i) \overset{C_t^{PV\_load}}{\ominus} Q_t^{TotalPV}(i) \qquad (16)$$

#### 2) Probabilistic ramp curve

According to the PRC definition and Eq. (13), the DPS of PRC $Q_t^{PRC}$ at period $t$ can be derived from the DDC between the DPS of net load at period $t + 1$ $Q_{t+1}^{PDC}$ and DPS of net load at period $t$ $Q_t^{PDC}$:

$$Q_t^{PRC}(i) = Q_{t+1}^{PDC}(i) \overset{C_t^{load t+1\_load t}}{\ominus} Q_t^{PDC}(i) \qquad (17)$$

### 2.3. Characteristic indices of the probabilistic duck curve and probabilistic ramp curve

#### 2.3.1. Indices

For the future application of the PDC and PRC in actual power system planning, four indices are designed, including the expected value curve, $\alpha\%$ confidence level curve, peak-to-valley difference, and probabilistic area. The indices quantify the most represented net load or ramp curve, uncertainty of net load and ramp, peak regulation demand, and PV curtailment, respectively.

#### 1) Expected value curve ($E_t^{netload}$ and $E_t^{ramp}$)

The expected value curves of net load and net load ramp denote their weighted average among all possible scenarios. The curves can be modeled as follows:

$$E_t^{netload} = \sum_{i=0} Q_t^{PDC}(i) \cdot (i\Delta p + P_t^{\min}) \qquad (18)$$

$$E_t^{ramp} = \sum_{i=0} Q_t^{PRC}(i) \cdot (i\Delta p + R_t^{\min}) \qquad (19)$$

where $P_t^{\min}$ and $R_t^{\min}$ denote the minimal net load and minimal net load ramp at period $t$, respectively. The expected value curve of the net load shows the averaged net load profile, which is usually the most representative net load scenario. The expected value curve of the ramp shows the average ramp capacity requirement at each period during one day.

#### 2) $\alpha\%$ confidence level curve ($CL_{\alpha\%}$)

This index is defined as the quantile of $(50 + \alpha/2)\%$ minus the quantile of $(50 - \alpha/2)\%$ and can be modeled as follows:

$$CL_{\alpha\%} = L_{(50+\alpha/2)\%} - L_{(50-\alpha/2)\%} \qquad (20)$$

where $L_{(50+\alpha/2)\%}$ and $L_{(50-\alpha/2)\%}$ denote the quantiles of $(50 + \alpha/2)\%$ and $(50 - \alpha/2)\%$ for the net load or net load ramp, respectively. The $\alpha\%$ confidence level curve quantifies the uncertainty of net load and net load ramp during one day.

3) Peak-to-valley difference (PTV, $Q^{PR}$)

This index is defined as the peak minus the valley value of the net load. It can be modeled as follows:

$$Q^{PR}(i) = Q^{PDC}_{peaktime}(i) \overset{C^{pk\_vy}}{\ominus} Q^{PDC}_{valleytime}(i) \qquad (21)$$

where $Q^{PDC}_{peaktime}$ and $Q^{PDC}_{valleytime}$ are the DPSs of the net load at the peak and valley times, respectively; and $C^{pk\_vy}$ is the dependence structure between the peak and valley times. Peak and valley times are derived from PDC expected value curve to avoid heavy computation due to peak and valley times uncertainty.

The PTV from the deterministic duck curve only reflects how much peak regulation is needed in an extreme or typical scenario. However, under high renewable energy penetration, both the peak regulation magnitude and its uncertainty should be considered. Therefore, the PTV from the PDC is necessary for a combination of different flexible resources to accommodate more PV generation.

4) Probabilistic area ($S$)

The PDC shows the extreme and average net loads with PV integration and exhibits the net load uncertainty from PV intermittency. In a power system with high PV penetration, it is usually not necessary to handle the extreme overgeneration scenario. The system should employ flexible resources that can cover most cases to balance economic operation and PV accommodation. The PDC provides a quantitative way to assess the capability of flexible resources to accommodate PV generation. We define the probabilistic region as the region between the minimal output of units (MOU) in the power system and the lower boundary of the PDC. The probabilistic area $S$ is defined as the area of the probabilistic region weighted by the net load probability. Thus the value of the probabilistic area represents the expected curtailment of PV generation during a day. The definition of the probabilistic area is demonstrated in Fig. 4.

According to the definition, the probabilistic area $S$ can be modeled as follows:

$$S = \sum_{t=t_{\min}}^{t_{\max}} \left( \sum_{i=0}^{I_t} Q^{PDC}_t(i)(i\Delta p + P^{\min}_t) \right) \qquad (22)$$

where $t$ denotes the time period; $i$ denotes the element serial number in $Q^{PDC}_t$; $P^{\min}_t$ denotes the minimal net load in period $t$; $t_{\min}$ and $t_{\max}$ are the periods corresponding to the intersection points between the MOU and minimal PDC curves (possible minimal net load curve); and $I_t$ is the maximal element serial number of $Q^{PDC}_t$ between the MOU and minimal PDC in period $t$, which can be calculated as follows:

$$I_t = \left[ \frac{P^{MOU}_t - P^{\min}_t}{\Delta p} \right] + 1 \qquad (23)$$

where [] is the floor function; $P^{MOU}_t$ denotes MOU at period $t$.

The marginal probabilistic area $\Delta S$ is defined as the marginal decrease of the probabilistic area under the unit decrement of the MOU. The marginal probabilistic area can quantify the marginal benefit of reducing the MOU measured by reducing PV curtailment and can be calculated as follows:

$$\Delta S = \frac{\Delta S'}{\Delta P} \qquad (24)$$

where $\Delta S'$ denotes the decrease of the probabilistic area when the MOU is reduced by $\Delta P$.

2.3.2. Application of indices

With the indices above, the PDC can be used to assist the flexible sources planning to accommodate PV generation. The expected value curves of net load and ramp approximately represent the most typical scenario and determine the basic requirements for peak regulation and flexible resources capacity. The PV accommodation depends on the capability of peak-to-valley regulation. The uncertainty of the PDC determines how frequently the flexible sources will be used to accommodate PV. The definition of the marginal probabilistic area quantitatively implies how much PV generation is expected to be accommodated by a unit of the extra flexible resource.

Taking the marginal probabilistic area for optimal MOU planning as an example, the marginal probabilistic area is used to quantify the marginal benefit of reducing the MOU measured by PV accommodation, and thus, $\Delta S$ will generally decrease when the MOU declines (shown with a green curve in Fig. 5). If we can estimate the marginal cost of reducing the MOU with a blue curve, then the optimal MOU could be approximately derived from the intersection point between the marginal benefit and marginal cost curves.

3. Empirical study of the Qinghai power system

In this section, an empirical study is conducted based on the Qinghai power system planning data in 2020. By then, the PV capacity in Qinghai will reach 10 GW, which is approximately 77% of the peak load (13 GW). The PV generation will meet over 20% of total load demand. The hydropower capacity will reach 15 GW, and the coal-fired unit capacity will be 5 GW. The purpose of this section is to validate the effectiveness of proposed method. Firstly, we model the distributions of PV generation and load considering their dependency. Then, the PDC and PRC of Qinghai are calculated and visualized. At the end of this section, the characteristics of the PDC and PRC are quantified and its application to power system flexible resource planning is presented.

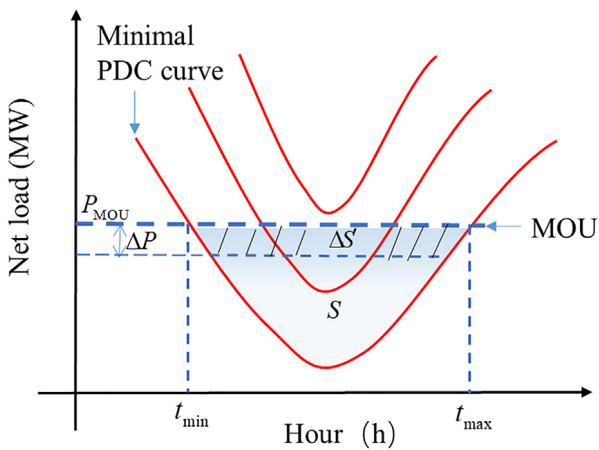

**Fig. 4.** Schematic diagram to calculate the probabilistic area (PV curtailment).

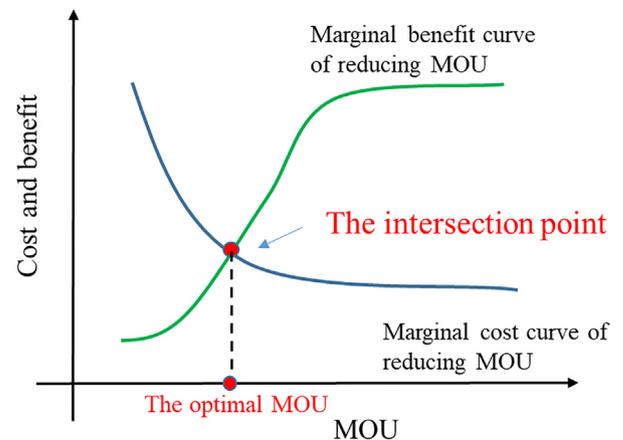

**Fig. 5.** Illustrative diagram of optimal MOU.

### 3.1. Marginal distribution modeling

Firstly, the marginal distribution of PV generation is modeled. Fig. 6 shows the generation distribution of a single PV farm in Qinghai at 12:00 PM. Performances of several fitting functions are compared, including kernel density function, beta function, normal function, Weibull function and real-data histogram. The real-data histogram has a long tail and high skewness value. The PV generation is mainly concentrated in the range from 0.6 p.u. to 0.9 p.u. Compared with other methods, the kernel density function achieves the best estimation to the probability distribution of PV generation.

### 3.2. Dependency modeling

The dependency between two PV farms in Golmud, Qinghai is modeled using the copula function, and their aggregated generation distribution is calculated using the DDC technique. Fig. 7 compares the aggregated generation distribution of the two PV farms with and without considering the Gaussian copula function. The statistic result from realistic data is set as the benchmark. As shown in Fig. 7, the red[1] curve considering the dependency is closer to the statistic distribution. Compared with the black curve without considering the dependency, the red curve considering the dependency successfully captures the information of fatter tail and lower peak value. This result occurs because the two PV farms with higher correlation tend to reach the maximal or minimal generation simultaneously, especially at noon. The results demonstrate the necessity of modeling the dependencies among PV farms in Qinghai.

Fig. 8 compares the distributions of the system net load at 12:00 PM with and without considering the dependency between total load and total PV generation. The results indicate that the correlation between total load and total PV generation is weak. This is because the industrial load accounts for more than 90% of the total load in Qinghai. The industry load is relatively smooth and less likely to be affected by weather.

Fig. 9 shows the distribution of system net load ramp at 4:00 PM when the sun starts going down and the demand for net load ramp capacity is considerable. The results imply that the dependency of net load between two adjacent periods is remarkable. Ignoring the dependency of net load between adjacent periods will introduce considerable error to the ramp distribution.

### 3.3. The probabilistic duck curve and probabilistic ramp curve of the Qinghai power system

Fig. 10 shows the PDC for the Qinghai power system in the spring of 2020. The PDC provides the expected value, uncertainty, and variability of net load demand. Variations in red shades represent the occurrence probability from 5% to 99%. The expectation of net load is shown by the black dotted curve. As shown in Fig. 10, due to the uncertainty of PV generation, the range of net load variation is wider during the daytime and thinner during the nighttime. The expectation of net load peaks at 12,000 MW at both 7:00 AM and 9:00 PM, the corresponding width of net load uncertainty is approximately 1500 MW. The minimum demand is approximately 6000 MW at 2:00 PM, whereas the corresponding width of net load uncertainty is more than 8000 MW. In the worst-case scenario, the net load valley reaches less than 2000 MW at 2:00 PM when nearly 80% of the load is supplied by PV generation. It should be noted that the occurrence probability of the worst-case scenario is less than 1%. This probability information cannot be observed from the deterministic duck curve.

---

[1] For interpretation of color in Fig. 7, the reader is referred to the web version of this article.

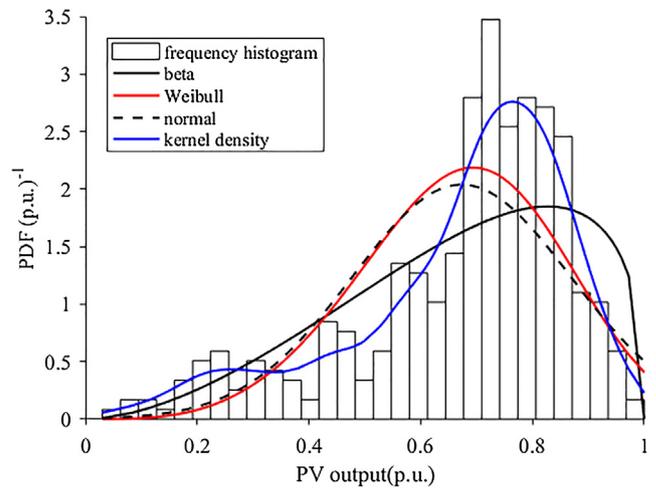

**Fig. 6.** Marginal distribution of a single PV generation at 12:00 PM in Qinghai.

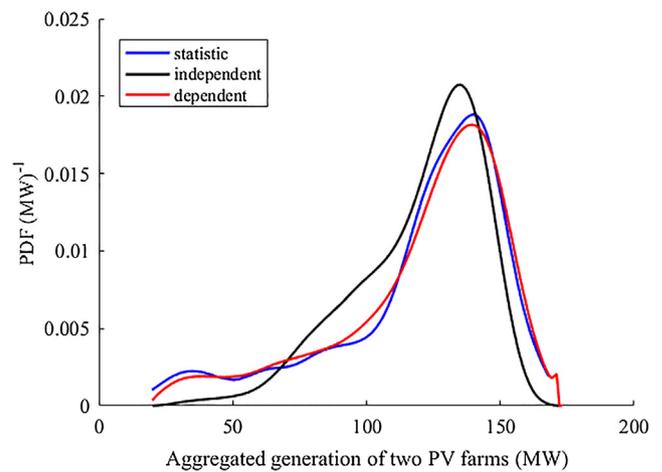

**Fig. 7.** The aggregated generation distribution of two PV farms in Golmud, Qinghai at 12:00 PM with three conditions: (1) considering the dependence structure between the generation of two PV farms, (3) without considering the dependence structure between the generation of two PV farms, and (3) the benchmark distribution directly from statistical data.

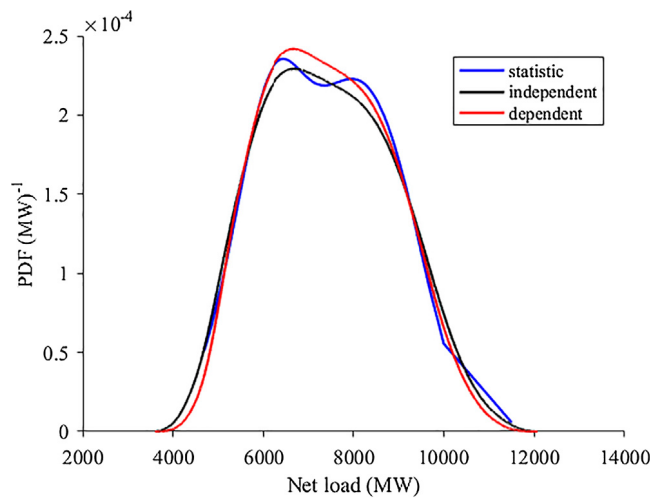

**Fig. 8.** Distribution of net load at 12:00 PM in Qinghai with three conditions: (1) considering the dependence structure between total load and total PV generation, (2) without considering the dependence structure between total load and total PV generation, and (3) the benchmark distribution directly from statistical data.

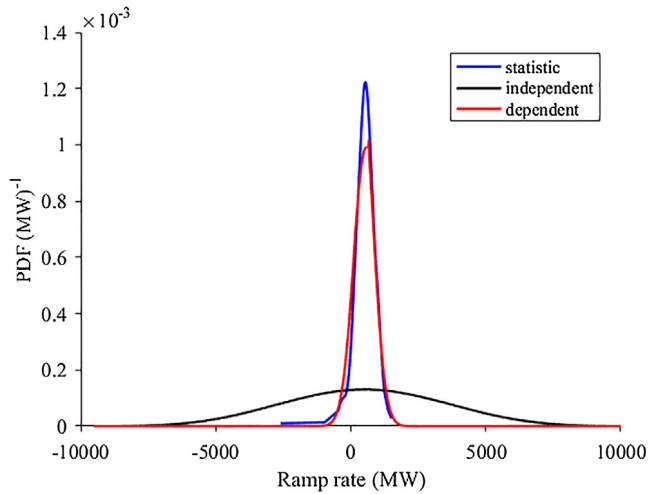

**Fig. 9.** Distribution of net load ramp at 4:00 PM with three conditions: (1) considering the dependence structure of net loads between the two adjacent periods, (2) without considering the dependence structure of net loads between the two adjacent periods, and (3) benchmark distribution directly from statistical data.

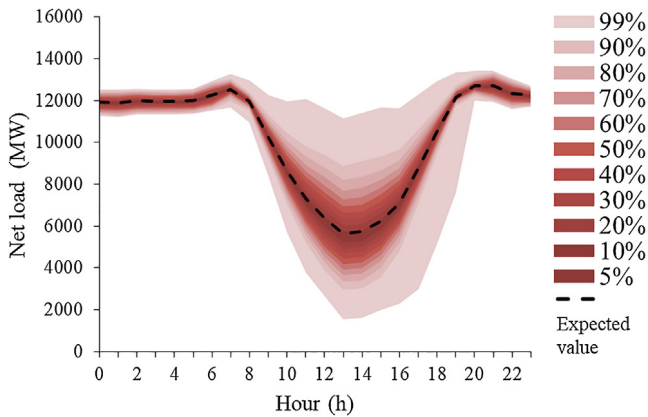

**Fig. 10.** The PDC of the Qinghai power system in the spring of 2020.

In the original load curve without PV generation, the demand valley happens in the early morning. The result of Fig. 10 indicates that high PV penetration tends to move the valley load period from early morning to noon with much larger uncertainty. Therefore, the scheduling of conventional generators and flexible resources would be fundamentally changed to participate in peak regulation and reserve provision. Those will significantly affect the cost efficiency of generators. Thus, this probabilistic net load profile should be considered in generation planning.

Fig. 11 shows the PRC for the Qinghai power system in the spring of 2020. The PRC provides the expected value, uncertainty, and variability of the net load ramp demand for each period of the day. As shown in Fig. 11, the ramp down demand and its uncertainty rapidly increase to the peak during the morning (from 6:00 AM to 8:00 AM), whereas the ramp up demand and its uncertainty reach their peak during dusk (from 4:00 PM to 6:00 PM). The maximum ramp down capacity demand appears at 8:00 AM with an expected value of approximately 2000 MW/h and a maximum value of approximately 3000 MW/h. The corresponding width of ramp down demand uncertainty is approximately 2000 MW/h. The maximum ramp up capacity demand appears at 6:00 PM with an expected value similar to the ramp down requirement (approximately 2000 MW/h). However, the corresponding ramp up demand uncertainty width is much larger (approximately 4000 MW/h). Due to the overlap of sunset and the coming of evening peak load, the highest possible ramp up requirement reaches approximately 4000 MW/h. Those results indicate that high PV penetrations increase the ramp requirement and its uncertainty. Flexible generators with excellent ramp performance are needed in the future generation mix. However, the utilization rate of such flexible resources may be low due to the increasing uncertainty of ramp requirement and low occurrence probability of extreme net load ramp. The strategic scheduling and planning of flexible resources are required.

In addition, Fig. 12 shows the index $CL_{99\%}$ (denote the index $CL_{\alpha\%}$ with $\alpha = 99$) of the PDC and PRC by black and red lines, respectively. The result suggests that $CL_{99\%}$ is able to directly quantify the PDC and PRC uncertainty. Obviously, large PDC uncertainty appears in the noon, and large PRC uncertainty appears at sunrise and sunset.

The PTV distribution is shown in Fig. 13. The PTV shows the flexible resource demand for peak regulation. A higher PTV implies a greater requirement for peak regulation resources. The PDF illustrates that the PTV is most likely in the range of 5000–9000 MW. The CDF shows that the probability of PTV greater than 5000 MW is approximately 0.9, and the PTV without subtracting PV generation is approximately 2000 MW. These results indicate that high PV penetration will enlarge the PTV nearly three times in 2020 in Qinghai.

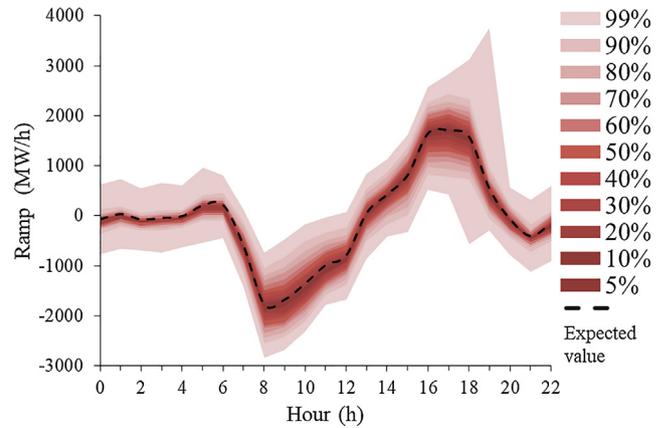

**Fig. 11.** The PRC of the Qinghai power system in the spring of 2020.

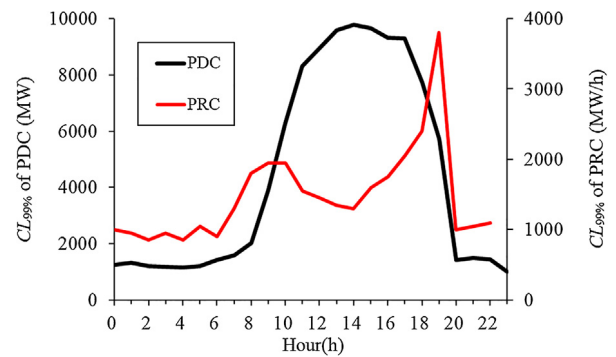

**Fig. 12.** $CL_{99\%}$ of PDC and PRC in Qinghai, 2020.

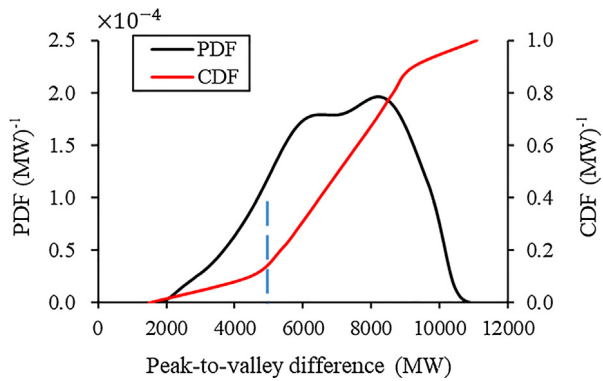

**Fig. 13.** PTV distribution in Qinghai, 2020.

### 3.4. Probabilistic duck curve for flexible resource planning

To further illustrate the application in actual power systems, the PDC (shown in Fig. 10) is used to analyze the planning of flexible resources such as coal-fired unit retrofitting, energy storage and demand response from a new technical-economic view.

It should be noted that this analysis is based on the following assumptions:

1) Qinghai is a hydropower-dominated power system. To better evaluate the impact of retrofitting on coal-fired units and generalize the method for other coal-dominated regions of China, a modified portfolio is performed, where the total coal power capacity is set to 10 GW, and hydropower capacity is set to 5 GW.
2) The maintenance cost change after coal-fired unit retrofitting can be ignored, compared with the fuel cost reduction and investment cost increase [28].
3) The deployment of energy storage cannot reduce the system minimal output directly but is able to charge using PV over-generation at noon and turn to discharge for the ramp up when PV generation decreases during dusk. Therefore, this approach is equivalent to reducing the minimal output from the perspective of accommodating PV and increasing system flexibility. Thus, it can be analyzed by the same method as the MOU proposed in Section 2.
4) Kumar et al. concluded that the greatest system benefit of coal-fired unit retrofitting comes from reducing the minimal output [21]. Thus, the retrofitting in this paper is referred to reducing the minimum output.
5) Due to the limitation of retrofitting technology, we assume that the minimal output of the coal-fired unit can be reduced to 30% of the installed capacity.

Fig. 14 shows the relationship among total PV curtailment, marginal PV curtailment, and system minimal output in Qinghai, 2020.

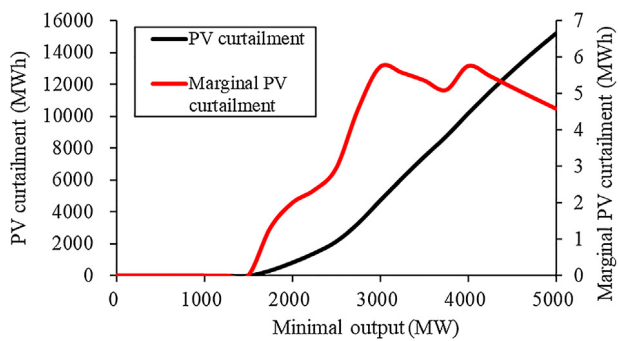

**Fig. 14.** Relationship among PV curtailment, marginal PV curtailment, and minimal output.

**Table 1**
Cost-benefit analysis of retrofitting a coal-fired unit and deployment of the energy storage system.

|  | Retrofit of coal-fired unit | Deployment of energy storage system (5 h) |
|---|---|---|
| Average investment cost USD/MW | 30,000 | 1,000,000 |
| Rate of annual revenue requirement | 16% | 16% |
| Annual investment cost USD/MW | 4800 | 160,000 |
| Daily investment cost USD/MW | 13.15 | 438.35 |
| Benefit of PV accommodation USD/MWh | 41.69 | 110.31 |
| Critical marginal cost represented by PV accommodation MWh/MW | 0.32 | 3.97 |

According to the total PV curtailment curve, if the minimal output rate of coal-fired units is reduced from 0.5 to 0.3, system minimal output will be reduced from 5000 MW to 3000 MW and the curtailment of PV will be decreased from 15,000 MWh to 5000 MWh each day. This result illustrates that coal-fired unit retrofitting has a remarkable performance in improving power system flexibility and accommodating renewable energy. The marginal PV curtailment curve shows that if the system minimal output is over 2700 MW, reducing minimal output by 1 MW can increase PV accommodation over 4 MWh each day.

We compare the cost-benefit of retrofitting coal-fired units and deploying energy storage. The results are summarized in Table 1. According to Ref. [31], the cost of reducing a coal-fired unit minimal output by 1 MW is approximately 30,000 USD. Therefore, the annual investment cost, which is the total cost multiplied by the rate of annual revenue requirement (RARR), is 4800 USD/MW, and the equivalent daily retrofitting cost is 13.15 USD/MW. From the perspective of maximizing social welfare, the benefit of accommodating renewable energy is quantified as the reducing fuel cost. Given the coal cost by 134.5 USD/t and the coal consumption rate by 0.31 t/MWh, the fuel cost of coal-fired units in China is approximately 41.69 USD/MWh. This can be regarded as the benefit of PV accommodation by retrofitting. Then, the break-even point of retrofitting coal-fired units is the retrofitting cost 13.15 USD/MW divided by the retrofitting benefit 41.69 USD/MWh, namely 0.32 MWh/MW. This means that if the marginal PV accommodation is greater than 0.32 MWh, the retrofit of a coal-fired unit is economical.

For the expensive energy storage, according to McKinsey, the cost is decreasing and could be 200 $/kWh in 2020 [32]. It is assumed that a 5-hour energy storage system is used in this paper, which means 1 MW energy storage is equipped with 5 MWh energy capacity. Thus, the daily cost of 1 MW energy storage is 438.35 USD with the RARR by 16%. The benefit of energy storage is not only the accommodation of renewable energy but also the provision of peak regulation capacity. In this paper, the benefit of providing peak regulation capacity could be approximately reflected by the difference between peak price and valley price, which is over 68.62 USD/MWh for commercial and industrial users in Qinghai [33]. Thus, the benefit of energy storage is the sum of PV accommodation benefit by 41.69 USD/MWh and the peak regulation benefit by 68.62 USD/MWh, namely 110.31 USD/MWh. Then, the break-even point of deploying energy storage is the cost 438.35 USD/MW divided by the benefit 110.31 USD/MWh, namely 3.97 MWh/MW. This means that if the marginal accommodation of PV is greater than 3.97 MWh, deploying energy storage is economical.

Fig. 15 shows the optimal planning capacity for flexible resources. Because the break-even point of retrofitting coal-fired units is much lower than that of deploying energy storage, we firstly determine the capacity of retrofitting coal-fired units. The break-even point of retrofitting coal-fired units is shown as point A in Fig. 15. The results indicate that the economic retrofitting plan can reduce the system minimal output down to 1500 MW. However, the system minimal output can only be reduced to as far as 3000 MW by retrofitting. Thus, the cross point would be point C, which implies that it would be economical to retrofit all available coal-fired capacity for reducing the system minimal output down-to 3000 MW and increasing PV accommodation by 10,000 MWh each day. To further reduce the system minimal output, expensive energy storage can be employed. The break-even point of deploying energy storage is shown as point B in Fig. 15. The results indicate that the system minimal output can be further reduced down to 2700 MW by deploying 1500 MWh (300 MW×5 h) of energy storage with increasing PV accommodation by 2000 MWh each day. For the minimal output interval from 1500 MW to 2700 MW, approaches with low capital cost such as demand response could be deployed, but this depends on the potential available resources on the demand side.

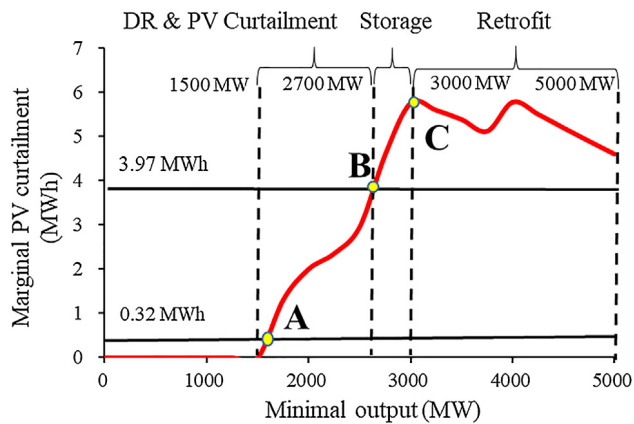

**Fig. 15.** Flexible resource planning using the PDC and marginal probabilistic area.

## 4. Discussion

The PDC and PRC provide useful information to guide power system generation planning for the Qinghai Province, China. For a power system with high PV penetration, adequate flexible resources should be planned to provide peak regulation and ramp capacity to accommodate intermittent PV generation. The peak regulation and ramp capacity should meet the net load shown by the PDC and PRC. The PDC and PRC are able to suggest the requirement for flexible generation capacity and how frequently the flexible generation capacity will be utilized. The large uncertain interval of the load and ramp indicates that it is uneconomical and impractical to install new flexible generation or expensive storage systems to meet the rare worst-case scenario.

Taking peak regulation in the Qinghai power system as an example, the PDC can be categorized into four sections (shown in Fig. 16). In the first section (valley load above 5000 MW), peak regulation is not a critical problem since the power system has the inherent flexibility (flexible generation schedule) to follow the load. In the second section (valley load from 3000 MW to 5000 MW), peak regulation capacity will be in shortage. Coal-fired unit retrofitting is suggested to enhance the power system flexibility since the investment cost for retrofitting and the operation cost for peak regulation are relatively low. In the third section (valley load from 2700 MW to 3000 MW), expensive energy storage is deployed to facilitate the PV accommodation while ensuring economic efficiency under the more severe situation. In the last section (valley load below 2700 MW), the occurrence probability is very low. Demand response and permitting PV curtailment will be the best choices, because they can be regarded as the flexible resources with nearly zero investment cost and high operation cost.

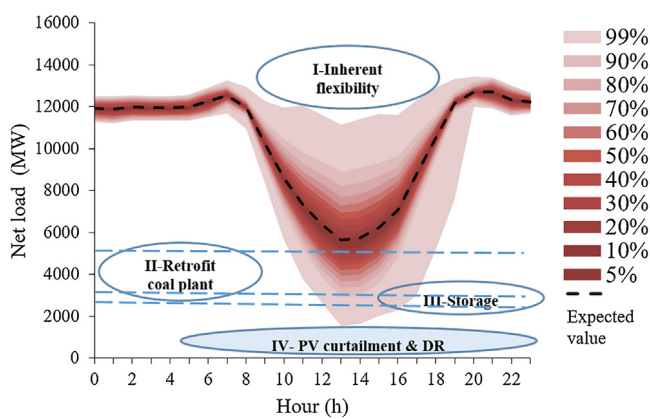

**Fig. 16.** Results of flexible resource planning in a modified Qinghai power system.

## 5. Conclusions and future works

In this paper, the concept of the duck curve is extended to the probabilistic duck curve and probabilistic ramp curve to describe the uncertainty of net load and net load ramp. The methodology of modeling PDC and PRC is proposed based on copula and DDC techniques, which are able to model the dependence structure among PV generation and loads. Several characteristic indices are designed to reflect the most represented scenario, uncertainty of net load and ramp, PV curtailment, and peak regulation demand. Further, an empirical study is conducted on the Qinghai power system for 2020 to validate the effectiveness of the proposed modeling method and to evaluate the technical-economics of flexible resources based on the PDC. The main conclusions are summarized as follows:

1) Kernel density estimation outperforms the parameter estimation function when modeling the marginal distribution of PV, and the copula function and DDC techniques are effective for modeling the dependence structure among PV generation and loads.
2) Integrating high PV penetration will move the valley period from early morning to noon and enlarge the Qinghai power system PTV nearly three times in 2020. Simultaneously, the minimal load faces major uncertainty driven by PV.
3) Integrating high PV penetration increases the average ramp capacity demand and its uncertainty. Therefore, larger flexible generators with higher ramp capacity are needed in the future generation mix; however, the utilization rate of such flexibility will be decreased due to the increase in the ramp capacity demand uncertainty.
4) The PDC and PRC are valuable for flexible resource planning. The study of minimal output with the proposed probability area shows that China's coal-fired units have great potential to improve the power system flexibility and accommodate renewable energy by retrofitting coal-fired units. However, due to the retrofitting capacity limitation of the coal-fired units, deploying a certain amount of more expensive energy storage can increase power system flexibility and accommodate more PV while also ensuring economic efficiency.
5) The PDC illustrates that the worst overgeneration scenarios rarely occur and that the deployment of flexible resources with heavy investment is uneconomical in these scenarios. Therefore, measures with nearly zero capital investment, such as demand response and curtailment of PV generation, can be regarded as flexible resources and should be adopted in some rare overgeneration scenarios to ensure economic planning and operation of the power system.

It should be noted that the proposed concepts, modeling method for the PDC and PRC, and the flexible resource planning method are all general. The kernel density estimation, copula function, and DDC do not rely on any unique condition in China. Therefore, it can be utilized in any power system with high PV penetration, such as the California power system, to guide power system planning.

Limited by the scope and space of this paper, the concepts and methodology of the PDC and PRC are especially focused on high PV penetration power systems and applied to flexible resource planning. Actually, the proposed concepts and modeling method could be further extended to the following:

1) Power system planning and operation, such as flexible planning with both PV and wind power, power system operation with the uncertainty of PV generation, wind generation, and load forecasting errors.
2) Other systems with uncertainty factors and dependence structure, such as multi-energy systems with energy price, generation and demand uncertainty.

Future works will focus on the analysis of flexible resource planning for power systems with both high wind power and PV penetration, and studying how the dependence structure among PV and wind farms helps accommodate renewable energy considering the wide geographical spread of China.

## Acknowledgments

This work is supported in part by the National Key Research and Development Program of China (No. 2016YFB0900100), National Natural Science Foundation of China (No. 51677096), Postdoctoral Innovative Talents Support Program (BX20180154), and Technical Projects of Qinghai Provincial Power Company of State Grid.

## References

[1] Obi M, Bass R. Trends and challenges of grid-connected photovoltaic systems – a review. Renew Sustain Energy Rev 2016;58:1082–94.
[2] Perlin J. From space to earth: the story of solar electricity. Cambridge, MA: Harvard University Press; 2000.
[3] National Energy Administration. Statistic data of PV generation in China; 2017, < http://www.nea.gov.cn/2018-01/24/c_136920159.htm > , 2018 [accessed 2018.06.01]. (in Chinese).
[4] John JS. Hawaii's solar-grid landscape and the 'nessie-curve'. Green Tech Media; 2014: < https://www.greentechmedia.com/articles/read/hawaiis-solar-grid-landscape-and-the-nessie-curve > . [accessed 2019.02.16].
[5] John JS. The california duck curve is real, and bigger than expected. Greentech media; 2016. < https://www. greentechmedia. com/articles/read/the-california-duck-curve-is-real-and-bigger-than-expected > . [accessed 2018.05.31].
[6] California independent system operator. What the duck curve tells us about managing a green grid; 2016, < https://www.caiso.com/Documents/FlexibleResourcesHelpRenewables_FastFacts.pdf > . [accessed 2018.05.31].
[7] Denholm P, O'Connell M, Brinkman G, Jorgenson J. Over-generation from solar energy in California: a field guide to the duck chart, National Renewable Energy Laboratory, Nov, 2015.Tech. Rep. NREL/TP-6A20-65023.
[8] Schoenung SM, Keller JO. Commercial potential for renewable hydrogen in California. Int J Hydrogen Energy 2017;42:13321–8.
[9] Zhang N, Lu X, McElroy MB, Nielsen CP, Chen X, Deng Y, et al. Reducing curtailment of wind electricity in China by employing electric boilers for heat and pumped hydro for energy storage. Appl Energy 2016;184:987–94.
[10] Denholm P, Margolis RM. Evaluating the limits of solar photovoltaics (PV) in electric power systems utilizing energy storage and other enabling technologies. Energy Policy 2007;35(9):4424–33.
[11] Komušanac I, Ćosić B, Duić N. Impact of high penetration of wind and solar PV generation on the country power system load: the case study of Croatia. Appl Energy 2016;184:1470–82.
[12] Chaudhary P, Rizwan M. Energy management supporting high penetration of solar photovoltaic generation for smart grid using solar forecasts and pumped hydro storage system. Renew Energy 2018;118:928–46.
[13] California independent system operator. Demand response and energy efficiency roadmap: maximizing preferred resources; 2013, < https://www.caiso.com/documents/dr-eeroadmap.pdf, 2013 > . [accessed 2018.05.31].
[14] Floch GL, Belletti F, Moura S. Optimal charging of electric vehicles for load shaping: a dual-splitting framework with explicit convergence bounds. IEEE Trans Transp Electrif 2016;2:190–9.
[15] Sanandaji BM, Vincent TL, Poolla K. Ramping rate flexibility of residential HVAC loads. IEEE Trans Sustain Energy 2016;7:865–74.
[16] Lazar J. Teaching the" duck" to fly. Regulatory assistance project; 2016.
[17] Hassan AS, Cipcigan L, Jenkins N. Optimal battery storage operation for PV systems with tariff incentives. Appl Energy 2017;203:422–41.
[18] Janko SA, Arnold MR, Johnson NG. Implications of high-penetration renewables for ratepayers and utilities in the residential solar photovoltaic (PV) market. Appl Energy 2016;180:37–51.
[19] Ding Y, Shao C, Yan J, Song Y, Zhang C, Guo C. Economical flexibility options for integrating fluctuating wind energy in power systems: the case of China. Appl Energy 2018;228:426–36.
[20] Sevilla FRS, Parra D, Wyrsch N, Patel M, Kienzle F, Korba P. Techno-economic analysis of battery storage and curtailment in a distribution grid with high PV penetration. J Storage Mater 2018;17:73–83.
[21] Kumar N, Venkataraman S, Lew D, Brinkman G, Palchak D, Cochran J. Retrofitting fossil power plants for increased flexibility. ASME 2014 Power Conference. American Society of Mechanical Engineers; 2014. V001T06A001-V001T06A001.
[22] Nelsen RB. An introduction to copulas. New York: Springer; 1999.
[23] Zhang N, Kang C, Singh C, Xia Q. Copula based dependent discrete convolution for power system uncertainty analysis. IEEE Trans Power Syst 2016;31:5204–5.
[24] Soleimanpour N, Mohammadi M. Probabilistic load flow by using nonparametric density estimators. IEEE Trans Power Syst 2013;28:3747–55.
[25] Ren Z, Yan W, Zhao X, Li W, Yu J. Chronological probability model of photovoltaic generation. IEEE Trans Power Syst 2014;29(3):1077–88.
[26] Sheather SJ, Jones MC. A reliable data-based bandwidth selection method for kernel density estimation. J R Statist Soc B 1991;53:683–90.
[27] Arbenz P. Bayesian copulae distributions, with application to operational risk management—some comments. Methodol Comput Appl Probability 2013;15(1):105–8. https://doi.org/10.1007/s11009-011-9224-0.
[28] Wikipedia. Kendall rank correlation coefficient; 2017. < https://en.wikipedia.org/wiki/Kendall_rank_correlation_coefficient > . [accessed 2018.05.31].
[29] Kang C, Xia Q, Xu W. Power system uncertainty analysis. Beijing, China: Science Press; 2011. (in Chinese).
[30] Wang Y, Zhang N, Kang C, Miao M, Shi R, Xia Q. An efficient approach to power system uncertainty analysis with high-dimensional dependencies. IEEE Trans Power Syst 2018;33:2984–94.
[31] Venkataraman S, Jordan G, O'Connor M, Kumar N, Lefton S, Lew D, et al. Cost-benefit analysis of flexibility retrofits for coal and gas-fueled power plants. Golden, Colorado: National Renewable Energy Laboratory (NREL); 2013.
[32] D'Aprile P, Newman J, Pinner D. The new economics of energy storage; 2016. < https://www.mckinsey.com/business-functions/sustainability-and-resource-productivity/our-insights/the-new-economics-of-energy-storage > . [accessed 2018.11.12].
[33] Qinghai Development and Reform Commission. Reducing electricity price for commercial and industrial users in Qinghai; 2018, < http://www.qhfgw.gov.cn/gzdt/tzgg/201809/t20180929_749115.shtml > . [accessed: 2018.11.12]. (in Chinese).